\documentstyle{article}
\begin{document}
\title{A Note on the scale symmetry \\
and Noether current}
\author{Naohisa OGAWA
\thanks{E-mail:~ogawa@particle.sci.hokudai.ac.jp, ~
s094305@math.sci.hokudai.ac.jp}
\\Department of Mathematics,  Hokkaido University
\\ Sapporo 060-0808 Japan}
\date{June,  1998}
\maketitle

\begin{abstract}
Usually we consider the symmetry of action as  the symmetry of the 
theory, however, in the Keplar problem the scaling symmetry existing in equation  of motion is not the ones for action.  It changes the multiplicative constant of action and the time boundary.  In such a case that the scale transformation does not leave the action  invariant but keeping the equation invariant, the following statement is proved. 
The time integration of Lagrangian is explicitly performed and the action can be expressed by the difference of formal (non-conserved) Noether charges at time boundaries.  In field theory the action can be expressed by the boundary integration of the formal Noether current.
\end{abstract}

\section{Scale invariance and Keplar motion}
The scale transformation of dynamical systems is widely used and discussed for a long time. 
The well known phenomena in classical mechanics is the Keplar's third theorem and the law of similitude for flow.  In the high energy physics the Bjorken's scaling law is well established fact \cite{Jackiw}, and further discussions depending on the renormalization group equations are essentially coming from the scale transformation.
The essential property of fractals is also the scale invariance. \\
~~~When the theory is called scale invariant, 
there are two different meanings. 
One is the invariance of equation of motion, another one is the invariance of action.  Even if the action changes its value by multiplicative constant, the equation of motion is invariant.  Therefore there are the transformations which make equation invariant and changes the action as discussed above \cite{Landau}. One example is the scale transformation of Keplar problem which will be seen later clearly.
Generally the invariance of equation holds when the invariance of action holds, but the opposite is not true. This is discussed in simple models by K.Cornelius \cite{Cor}.
In this paper we discuss the scale transformation of dynamical systems, where the equation is invariant but not the action. We show in such a situation the action becomes the functional on the boundary similar to the topological field theory.\\

~~~Keplar motion is understood by the equation as
\begin{equation}
\frac{d^2 {\bf r}}{d t^2} = - \frac{k{\bf r}}{r^3},
\end{equation}
which is invariant under the scale transformation:
\begin{equation}
{\bf r} \to {\bf r'}= \alpha{\bf r},~~~t \to t'= \alpha^{3/2}t.
\end{equation}
The related Lie transformation is given by
\begin{equation}
{\bf r}(t) \to {\bf r'}(t) = \alpha{\bf r}(\alpha^{-3/2}t).
\end{equation}
Both keep the equation invariant, which is well known 
as the Keplar's third theorem.
However, the action
\begin{equation}
S(T_f,T_i) = \int^{T_f}_{T_i} dt ~[ \frac{1}{2} (\frac{d {\bf r}}{dt})^2 
+  \frac{k}{r}~],
\end{equation}
changes into the form:
\begin{equation}
S'(T_f,T_i) = \alpha^{1/2} S(\alpha^{-3/2}T_f, \alpha^{-3/2}T_i).
\end{equation}
Therefore the scale symmetry does not hold in the action.
This is also known as the ``Dynamical similitude" discussed by Landau and Lifshitz \cite{Landau}. It should be stressed that this kind of phenomenon also occurs in type IIB string theory \cite{Koch}.
For one equation of motion, corresponding action is not unique.
The additional degree of freedom of action giving the same 
equation of motion is the multiplicative constant and boundary. 
This explains our result for Keplar problem.
Therefore we can generally suppose the following point.
If we have the invariant equation of motion under the scale 
transformation:
\begin{equation}
q^i(t) \to q'^i(t) \equiv \alpha q^i(\alpha^{-d} t),
\end{equation}
Then the related action should be changed in the following way:
\begin{equation}
S'(T_f,T_i) = \alpha^{2-d} S(\alpha^{-d}T_f, \alpha^{-d}T_i).
\end{equation}
This is because the action has dimension $[L]^2 [T]^{-1}$.

\section{General scale invariant theory}
Now we consider the formal Noether charge corresponding to the infinitesimal transformation 
$\alpha = 1 + \epsilon$.
\begin{eqnarray}
\delta S \equiv S'(T_f,T_i)-S(T_f,T_i) &=& 
\int^{T_f}_{T_i}dt [\delta q^i (\frac{\partial L}{\partial q^i} - 
\frac{d}{dt}\frac{\partial L}{\partial {\dot q}^i}) + 
\frac{d}{dt}(\frac{\partial L}{\partial {\dot q}^i} \delta q^i)] \nonumber\\
&=& \epsilon [Q(T_f)-Q(T_i)],
\end{eqnarray}
where $Q$ is the Noether charge defined by
\begin{equation}
Q \equiv \frac{\partial L}{\partial {\dot q}^i} 
\frac{\delta q^i}{\epsilon}.
\end{equation}
On the other hand, $\delta S$ can be expressed as follows.
\begin{eqnarray}
\delta S &\equiv& S'(T_f,T_i)-S(T_f,T_i) = \alpha^{2-d} 
S(\alpha^{-d}T_f, \alpha^{-d}T_i) - S(T_f,T_i) \nonumber\\
&=& (1+(2-d)\epsilon) S((1-d\epsilon)T_f, (1-d\epsilon)T_i) 
-S(T_f,T_i)\nonumber\\
 &= & (2-d)\epsilon S(T_f,T_i) -d\epsilon[T_f L(T_f) - T_i L(T_i)].
\end{eqnarray}
Therefore we obtain the equality
\begin{equation}
(2-d) S(T_f,T_i) = [Q(T_f) + d T_f L(T_f)] - [Q(T_i) +d T_i L(T_i)] .
\end{equation}
If the action is invariant under scale transformation up to the boundary (d=2),
the conserved charge is  
\begin{equation}
\tilde{Q}(t) \equiv Q(t) + d t L(t).
\end{equation}
Using this replaced Noether charge, our result is
\begin{equation}
(2-d) S(T_f,T_i) = \tilde{Q}(T_f) -  \tilde{Q}(T_i).
\end{equation}
In the case of Keplar motion, $d=3/2$ and 
$$\tilde{Q} = {\bf r} \cdot \dot{\bf r} -\frac{3}{2}t{\dot {\bf  r}}^2 + \frac{3}{2}t(\frac{1}{2}{\dot {\bf  r}}^2 + \frac{k}{r})$$
is the charge.  Then we have
\begin{equation}
S = [\frac{d {\bf r}^2}{dt} ]^{T_f}_{T_i} - 3E(T_f - T_i),
\end{equation}
where E is the total energy. 
In this way the action becomes the difference of 
formal Noether charges at time boundaries.

\section{Field Theory}
The extension to the field theory is straight forward.
Let us work with the action in D-dimensional space time.
\begin{equation}
S = ~\int_{\Omega} d^Dx ~
[\frac{1}{2} \partial_{\mu} \Phi \partial^{\mu} \Phi 
~- ~V(\Phi)~],
\end{equation}
while the equation of motion takes the form as
\begin{equation}
 \partial^2 \Phi + \partial_\Phi V(\Phi) =0.
\end{equation}
Suppose that the equation of motion is invariant under the 
scale (Lie) transformation
\begin{equation}
\Phi(x) ~\to~  \Phi'(x) \equiv \alpha \Phi(\alpha^{-d} x).
\end{equation}
Then we have relation
$$ V(\Phi) \to \alpha^{2-2d}V(\Phi),$$
which leads to the change of Lagrangian density
$$ L(\Phi(x)) \to \alpha^{2-2d} L(\Phi(\alpha^{-d}x)).$$
So we obtain
\begin{equation}
S'[\Omega] = \alpha^{2+(D-2)d} S[\Omega'],~~~
\Omega' = \alpha^{-d} \Omega.
\end{equation}
Then we have the relation as we have done before,
\begin{eqnarray}
\delta S[\Omega] &=& S'-S = (1+\{ 2+(D-2)d \} \epsilon)(S[\Omega] - 
\epsilon \int_{\Omega} d^Dx \partial_\mu (d L x^{\mu})) - 
S[\Omega], \nonumber\\
&=& \epsilon [\{2+(D-2)d \} S[\Omega] - \int_{\Omega} d^Dx \partial_\mu (d L x^{\mu})].
\end{eqnarray}
And has another form
\begin{eqnarray}
\delta S[\Omega] &=&  \int_{\Omega}d^Dx [\delta \Phi (\frac{\partial L}{\partial \Phi} - 
\frac{\partial}{\partial x^\mu} 
\frac{\partial L}{\partial \Phi_{,\mu}}) + 
\frac{\partial}{\partial x^\mu}(\frac{\partial L}{\partial \Phi_{,\mu}} \delta \Phi)] \nonumber\\
&=& \epsilon \int_{\Omega}d^Dx \partial_\mu J^{\mu}(x),
\end{eqnarray}
where $J^\mu \equiv \frac{\partial L}{\partial \Phi_{,\mu}} 
\frac{\delta \Phi}{\epsilon}. $
At last, we come to the final form
\begin{equation}
\{2+(D-2)d \} S[\Omega] = 
\int_{\Omega} d^Dx \partial_\mu \tilde{J}^\mu  = 
\tilde{Q}(T_f) -\tilde{Q}(T_i),~~~
\tilde{Q} \equiv \int d^{D-1}x ~\tilde{J}^0(x).
\end{equation}
The last equality holds when the spatial boundary is negligible.
The redefined Noether current is defined by
\begin{eqnarray}
\tilde{J}^\mu &=&\frac{\partial L}{\partial \Phi_{,\mu}} \frac{\delta \Phi}{\epsilon} + d L x^{\mu}, 
\nonumber \\
&=& \frac{\partial L}{\partial \Phi_{,\mu}} (\Phi-d\epsilon x^\mu \partial_\mu \Phi) + dLx^\mu, 
\nonumber \\
&=& \frac{\partial L}{\partial \Phi_{,\mu}} \tilde{\delta}\Phi - T^{\mu}_{~\nu} \tilde{\delta} x^\nu,
\end{eqnarray}
where $\tilde{\delta}$ is the infinitesimal transformation
$$ \tilde{\delta}\Phi \equiv [\alpha \Phi - \Phi ]/\epsilon = \Phi, ~~~ \tilde{\delta} x^\mu \equiv [\alpha^d x^\mu- x^\mu]/\epsilon = dx^\mu,$$
and $T^{\mu \nu}$ is the well known canonical energy momentum tensor defined by
$$T^{\mu}_{~\nu} \equiv \frac{\partial L}{\partial \Phi_{,\mu}} \partial_\nu \Phi - L \delta^\mu_\nu.$$
Therefore we again obtained the relation that for the scale transformation which makes equation of motion invariant but changes action,  its action equals to the boundary integration of formal (non-conserved) Noether current.

\section{Discussion}
We have shown the action becomes the boundary integration of formal (non-conserved) Noether current when the scale transformation makes equation invariant but changes its action.
Our program holds in many scale invariant (equation) theories.
But our method depends on the on-shell condition, that is, 
our boundary valued action is essentially classical. 
So we can not use it for the full quantum theory, 
but may be applied for the WKB method and the 
theory of instantons. 
The application to the type IIB string theory is also the interesting 
target, where the rescaling symmetry changes the action 
with setting the equation of motion invariant.
These are still open problems.
\\

\noindent{\em Acknowledgment}\\
The author would like to thank Prof.T.Okazaki for discussion and encouraging him. He also thanks Prof.R.M.Koch for giving him the interesting information on string theory. He especially thanks Prof.R.Jackiw for the interest and important information of related topics.

\end{document}